\documentclass[table, a4paper,11pt]{article}
\usepackage{pos}

\usepackage{makecell}
\usepackage{color}
\usepackage{xcolor}
\usepackage{diagbox}
\usepackage{blkarray}

\allowdisplaybreaks[1]
\newcommand{\Op}{\mathcal{O}}
\newcommand{\Opr}{\mathcal{R}}

\newcommand{\tr}[1]{{\rm Tr}\left(#1\right)}

\newcommand{\rmi}{{\rm i}}

\newcommand{\befun}{ \dot{C}}

\title{Renormalization Group Equations for the Dimension-7 SMEFT Operators}
\author*[a]{Di Zhang}
\affiliation[a]{Department of Physics, Technical University of Munich,\\
	James-Franck-Stra\ss e, 85748 Garching, Germany}

\emailAdd{di1.zhang@tum.de}
\abstract{
	In this talk, a Green's basis and a new physical basis for dimension-seven (dim-7) operators in the Standard Model effective field theory (SMEFT) are proposed. The reduction relations between those two bases are also presented, where some redundant dim-6 operators in the Green's basis are involved if the dim-5 operator exists. Taking advantage of these two bases for dim-7 operators and the associated reduction relations, we work out the complete one-loop renormalization group equations (RGEs) for dim-5 and dim-7 operators up to $\mathcal{O} \left( \Lambda^{-3} \right)$ in the SMEFT, including not only the mixing among the same dimensional operators but also that among different dimensional ones. These results can be exploited to study full one-loop RG effects on some lepton- or baryon-number-violating processes up to $\mathcal{O} \left( \Lambda^{-3} \right)$ in the SMEFT, such as neutrino masses, neutrinoless double beta decay, meson and nucleon decays.
}

\FullConference{42nd International Conference on High Energy Physics (ICHEP2024)\\
18-24 July 2024\\
Prague, Czech Republic\\}

\begin{document}
\maketitle

\section{Introduction}

The Standard Model (SM) of particle physics is very successful in describing strong, weak and electromagnetic interactions and has passed a plethora of precision tests~\cite{ParticleDataGroup:2024cfk}, but it is unable to explain neutrino masses, dark matter and the matter-antimatter asymmetry of the Universe, and hence believed to be incomplete~\cite{ParticleDataGroup:2024cfk,Xing:2020ijf}. In the case where the energy scale $\Lambda$ of new physics is much higher than the electroweak scale, one can make use of the SM effective field theory (SMEFT)~\cite{Buchmuller:1985jz,Grzadkowski:2010es} 
\begin{eqnarray}
\mathcal{L}^{}_{\rm SMEFT} = \mathcal{L}^{}_{\rm SM} + \frac{1}{2} \left( C^{\alpha\beta}_5 \Op^{(5)}_{\alpha\beta} + {\rm h.c.} \right) + \sum_i C^i_6 \Op^{(6)}_i + \sum_j C^j_7 \Op^{(7)}_j + \dots \;,
\end{eqnarray}
to model-independently describe and study indirect low-energy consequences of new physics that are encoded in Wilson coefficients $C^{}_d$ of non-renormalizable operators $\Op^{(d)}$ with $d>4$ being the mass dimension. In the SMEFT, operators $\Op^{(d)}$ consist of the SM fields and preserve the SM gauge symmetry and Lorentz invariance, and their Wilson coefficients $C^{}_d$ are suppressed by $1/\Lambda^{d-4}$. 

To investigate low-energy consequences of new physics within the SMEFT, one has to construct an operator basis that contains a set of independent operators. The number of independent operators at each mass dimension is definite and can be worked out by the Hilbert series~\cite{Henning:2015alf}, but the specific operators in a physical basis are usually ambiguous~\cite{Brivio:2017vri,Isidori:2023pyp}. The dim-5 operator is unique and known as the Weinberg operator~\cite{Weinberg:1979sa}. For dim-6 operators, the most popular basis is the so-called Warsaw basis~\cite{Grzadkowski:2010es}. The physical basis for dim-7 operators has been constructed in Refs.~\cite{Lehman:2014jma,Liao:2016hru,Liao:2019tep} (see the latest review~\cite{Isidori:2023pyp} and references therein for the higher-dimensional-operator basis). In general, those operators are generated at the cut-off scale $\Lambda$ where new physics decouples, and their Wilson coefficients are obtained via matching conditions at $\Lambda$. Then, one may make use of the SMEFT renormalization group equations (RGEs) to run those Wilson coefficients from the cut-off scale $\Lambda$ down to the electroweak scale so as to encounter them with precision observables~\cite{Henning:2014wua}. Due to the RG mixing among different operators, an operator not generated by matching can achieve a non-vanishing Wilson coefficient through running. Such non-trivial relations among operators need to be carefully considered in a phenomenological analysis~\cite{Isidori:2023pyp}. Therefore, the SMEFT RGEs have been extensively discussed in literature, e.g., see Ref.~\cite{Isidori:2023pyp} and references therein. 

In this talk, we focus on RGEs of dim-5 and dim-7 operators up to $\mathcal{O} \left( \Lambda^{-3} \right) $, whose forms are 
\begin{eqnarray}\label{eq:rge-form}
	\befun^{}_5 &=& \gamma^{(5,5)} C^{}_5 + \hat{\gamma}^{(5,5)} C^{}_5 C^{}_5 C^{}_5+ \gamma^{(5,6)}_i C^{}_5 C^{i}_6 + \gamma^{(5,7)}_i C^{i}_7 \;,
	\nonumber
	\\
	\befun^{i}_7 &=& \gamma^{(7,7)}_{ij} C^{j}_7 + \gamma^{(7,5)}_i C^{}_5 C^{}_5 C^{}_5 + \gamma^{(7,6)}_{ij} C^{}_5 C^{j}_6 \;
\end{eqnarray}
with $\befun^{\cdots}_{\cdots} \equiv \mu {\rm d} C^{\cdots}_{\cdots} / {\rm d} \mu$ and $\gamma^{\dots}_{\dots}$ being the anomalous dimension matrix. $\gamma^{(5,5)}$ and $\gamma^{(7,7)}$ in Eq.~\eqref{eq:rge-form} have been taken into consideration in Refs.~\cite{Babu:1993qv,Chankowski:1993tx,Antusch:2001ck} and~\cite{Liao:2016hru,Liao:2019tep}, respectively, and others have been partially discussed with some approximations in Ref.~\cite{Chala:2021juk}. We attempt to complete the RGEs for all dim-5 and dim-7 operators up to $\mathcal{O} \left( \Lambda^{-3} \right)$ without any approximation. To achieve our goal, we first construct a new physical basis for dim-7 operators, which is more suitable for calculations and to get compact results compared with the one proposed in Ref.~\cite{Liao:2019tep}. Moreover, the so-called Green's basis~\cite{Jiang:2018pbd} that is directly related to 1PI Green's functions and needed in intermediate calculations with off-shell scheme is also constructed for dim-7 operators, together with the reduction relations converting the Green's basis to the physical one. With those two bases and the reduction relations between them, we derive the complete one-loop RGEs of dim-5 and dim-7 operators up to $\mathcal{O} \left( \Lambda^{-3} \right) $.

\section{Operator Bases for Dim-7 Operators}

\begin{table}[t!]
	\centering
	\renewcommand\arraystretch{1}
	\resizebox{0.9\textwidth}{!}{
		\begin{tabular}{l|c|l|c}
			\hline\hline
			\multicolumn{2}{c|}{$\psi^2 H^4$} & \multicolumn{2}{c}{$\psi^4 H$} \\
			\hline
			$\Op^{(S)\alpha\beta}_{\ell H}$ & \makecell[c]{ $\displaystyle\frac{1}{2} \left( \Op^{\alpha\beta}_{\ell H} + \Op^{\beta\alpha}_{\ell H} \right) $ \\ with $\Op^{\alpha\beta}_{\ell H} = \epsilon^{ab} \epsilon^{de}  \left( \ell^a_{\alpha\rm L} C \ell^d_{\beta\rm L} \right) H^b H^e \left( H^\dagger H \right)$ } & $\Op^{(S)\alpha\beta\gamma\lambda}_{\overline{e} \ell\ell\ell H}$ & \makecell[c]{ $\displaystyle\frac{1}{6} \left( \Op^{\alpha\beta\gamma\lambda}_{\overline{e} \ell\ell\ell H} + \Op^{\alpha\lambda\beta\gamma}_{\overline{e} \ell\ell\ell H}  + \Op^{\alpha\gamma\lambda\beta}_{\overline{e} \ell\ell\ell H}  + \Op^{\alpha\beta\lambda\gamma}_{\overline{e} \ell\ell\ell H}  + \Op^{\alpha\gamma\beta\lambda}_{\overline{e} \ell\ell\ell H}  + \Op^{\alpha\lambda\gamma\beta}_{\overline{e} \ell\ell\ell H} \right) $ \\ with $\Op^{\alpha\beta\gamma\lambda}_{\overline{e} \ell\ell\ell H} = \epsilon^{ab} \epsilon^{de} \left( \overline{E^{}_{\alpha\rm R}} \ell^a_{\beta\rm L}\right) \left( \ell^b_{\gamma\rm L} C \ell^d_{\lambda\rm L} \right) H^e$}  \\
			\cline{1-2}
			\multicolumn{2}{c|}{$\psi^2 H^3 D$} & $\Op^{(A)\alpha\beta\gamma\lambda}_{\overline{e} \ell\ell\ell H}$ & $\displaystyle\frac{1}{6} \left( \Op^{\alpha\beta\gamma\lambda}_{\overline{e} \ell\ell\ell H} + \Op^{\alpha\lambda\beta\gamma}_{\overline{e} \ell\ell\ell H}  + \Op^{\alpha\gamma\lambda\beta}_{\overline{e} \ell\ell\ell H}  - \Op^{\alpha\beta\lambda\gamma}_{\overline{e} \ell\ell\ell H}  - \Op^{\alpha\gamma\beta\lambda}_{\overline{e} \ell\ell\ell H}  - \Op^{\alpha\lambda\gamma\beta}_{\overline{e} \ell\ell\ell H} \right) $ \\
			\cline{1-2}
			$\Op^{\alpha\beta}_{\ell e H D}$ & $\epsilon^{ab} \epsilon^{de} \left( \ell^a_{\alpha\rm L} C \gamma^{}_\mu E^{}_{\beta\rm R} \right) H^b H^d {\rm i} D^\mu H^e$ & $\Op^{(M)\alpha\beta\gamma\lambda}_{\overline{e} \ell\ell\ell H}$ &  $\displaystyle\frac{1}{3} \left( \Op^{\alpha\beta\gamma\lambda}_{\overline{e} \ell\ell\ell H} +  \Op^{\alpha\gamma\beta\lambda}_{\overline{e} \ell\ell\ell H} - \Op^{\alpha\lambda\gamma\beta}_{\overline{e} \ell\ell\ell H} - \Op^{\alpha\gamma\lambda\beta}_{\overline{e} \ell\ell\ell H} \right) $ \\
			\cline{1-2}
			\multicolumn{2}{c|}{$\psi^2 H^2 D^2$} & $\Op^{\alpha\beta\gamma\lambda}_{\overline{d} \ell q \ell H 1}$ & $\epsilon^{ab} \epsilon^{de} \left( \overline{D^{}_{\alpha\rm R}} \ell^a_{\beta\rm L} \right) \left( Q^b_{\gamma\rm L} C \ell^d_{\lambda\rm L} \right) H^e $  \\
			\cline{1-2}
			$ \Op^{(S)\alpha\beta}_{\ell HD1} $ & \makecell{ $ \displaystyle\frac{1}{2} \left( \Op^{\alpha\beta}_{\ell HD1} + \Op^{\beta\alpha}_{\ell HD1} \right) $  \\ with $\Op^{\alpha\beta}_{\ell HD1} = \epsilon^{ab} \epsilon^{de} \left( \ell^a_{\alpha\rm L} C D^\mu \ell^b_{\beta\rm L} \right) H^d D^{}_\mu H^e$}  & $\Op^{\alpha\beta\gamma\lambda}_{\overline{d} \ell q \ell H 2}$ & $\epsilon^{ad} \epsilon^{be} \left( \overline{D^{}_{\alpha\rm R}} \ell^a_{\beta\rm L} \right) \left( Q^b_{\gamma\rm L} C \ell^d_{\lambda\rm L} \right) H^e$  \\
			$ \Op^{(S)\alpha\beta}_{\ell HD2}  $ & \makecell{ $ \displaystyle\frac{1}{2} \left( \Op^{\alpha\beta}_{\ell HD2} + \Op^{\beta\alpha}_{\ell HD2} \right)$ \\ with $\Op^{\alpha\beta}_{\ell HD2} = \epsilon^{ad} \epsilon^{be} \left( \ell^a_{\alpha\rm L} C D^\mu \ell^b_{\beta\rm L} \right) H^d D^{}_\mu H^e$ } & $\Op^{\alpha\beta\gamma\lambda}_{\overline{d} \ell u e H}$ & $\epsilon^{ab} \left( \overline{D^{}_{\alpha\rm R}} \ell^a_{\beta\rm L} \right) \left( U^{}_{\gamma\rm R} C E^{}_{\lambda\rm R} \right) H^b$ \\
			\cline{1-2}
			\multicolumn{2}{c|}{$\psi^2 H^2 X$} & $\Op^{\alpha\beta\gamma\lambda}_{\overline{q} u \ell \ell H}$ & $\epsilon^{ab} \left( \overline{Q^{}_{\alpha\rm L}} U^{}_{\beta\rm R} \right) \left( \ell^{}_{\gamma\rm L} C \ell^a_{\lambda\rm L} \right) H^b$  \\
			\cline{1-2}
			$\Op^{(A)\alpha\beta}_{\ell HB}$ & \makecell[c]{ $\displaystyle \frac{1}{2} \left( \Op^{\alpha\beta}_{\ell HB} - \Op^{\beta\alpha}_{\ell HB} \right)$ \\ with $\Op^{}_{\ell HB} = \epsilon^{ab} \epsilon^{de} \left( \ell^a_{\alpha\rm L} C \sigma^{}_{\mu\nu} \ell^d_{\beta\rm L} \right) H^b H^e B^{\mu\nu}$} & $\Op^{(M)\alpha\beta\gamma\lambda}_{\overline{\ell}dddH}$  & \makecell[c]{ $\displaystyle\frac{1}{3} \left( \Op^{\alpha\beta\gamma\lambda}_{\overline{\ell}dddH} + \Op^{\alpha\gamma\beta\lambda}_{\overline{\ell}dddH} - \Op^{\alpha\beta\lambda\gamma}_{\overline{\ell}dddH} - \Op^{\alpha\lambda\beta\gamma}_{\overline{\ell}dddH} \right) $ \\ with $\Op^{\alpha\beta\gamma\lambda}_{\overline{\ell}dddH} = \left( \overline{\ell^{}_{\alpha\rm L}} D^{}_{\beta\rm R} \right) \left( D^{}_{\gamma\rm R} C D^{}_{\lambda\rm R} \right) H$}   \\
			$\Op^{\alpha\beta}_{\ell HW}$ & $\epsilon^{ab} \left( \epsilon \sigma^I \right)^{de} \left( \ell^a_{\alpha\rm L} C \sigma^{}_{\mu\nu} \ell^d_{\beta\rm L} \right) H^b H^e W^{I \mu\nu}$ & $\Op^{\alpha\beta\gamma\lambda}_{\overline{\ell} dud \widetilde{H}}$ & $\left( \overline{\ell^{}_{\alpha\rm L}} D^{}_{\beta\rm R} \right) \left( U^{}_{\gamma\rm R} C D^{}_{\lambda\rm R} \right) \widetilde{H} $  \\
			\cline{1-2}
			\multicolumn{2}{c|}{$\psi^4 D$} &  $\Op^{\alpha\beta\gamma\lambda}_{\overline{\ell} d qq \widetilde{H}}$ & $\epsilon^{ab} \left( \overline{\ell^{}_{\alpha\rm L}} D^{}_{\beta\rm R} \right) \left( Q^{}_{\gamma\rm L} C Q^a_{\lambda\rm L} \right) \widetilde{H}^b$ \\
			\cline{1-2}
			$\Op^{(S)\alpha\beta\gamma\lambda}_{\overline{e} dddD} $   &   \makecell[c]{$ \displaystyle\frac{1}{6} \left( \Op^{\alpha\beta\gamma\lambda}_{\overline{e}dddD} + \Op^{\alpha\lambda\beta\gamma}_{\overline{e}dddD} +  \Op^{\alpha\gamma\lambda\beta}_{\overline{e}dddD} + \Op^{\alpha\beta\lambda\gamma}_{\overline{e}dddD} + \Op^{\alpha\gamma\beta\lambda}_{\overline{e}dddD} + \Op^{\alpha\lambda\gamma\beta}_{\overline{e}dddD} \right)$  \\ with $\Op^{\alpha\beta\gamma\lambda}_{\overline{e}dddD} = \left(\overline{E^{}_{\alpha\rm R}} \gamma^{}_\mu D^{}_{\beta\rm R} \right) \left( D^{}_{\gamma\rm R} C {\rm i} D^\mu D^{}_{\lambda\rm R} \right) $} &  $\Op^{(A)\alpha\beta\gamma\lambda}_{\overline{e}qdd\widetilde{H}}$ & \makecell[c]{ $\displaystyle\frac{1}{2} \left( \Op^{\alpha\beta\gamma\lambda}_{\overline{e}qdd\widetilde{H}} - \Op^{\alpha\beta\lambda\gamma}_{\overline{e}qdd\widetilde{H}} \right) $ \\ with $\Op^{\alpha\beta\gamma\lambda}_{\overline{e}qdd\widetilde{H}} = \epsilon^{ab} \left( \overline{ E^{}_{\alpha\rm R}} Q^a_{\beta\rm L} \right) \left( D^{}_{\gamma\rm R} C D^{}_{\lambda\rm R} \right) \widetilde{H}^b$} \\
			$\Op^{(S)\alpha\beta\gamma\lambda}_{\overline{d}u\ell\ell D}$ &   \makecell[c]{$ \displaystyle\frac{1}{2} \left( \Op^{\alpha\beta\gamma\lambda}_{\overline{d} u\ell\ell D} + \Op^{\alpha\beta\lambda\gamma}_{\overline{d} u\ell\ell D} \right) $ \\ with $\Op^{\alpha\beta\gamma\lambda}_{\overline{d} u\ell\ell D} = \epsilon^{ab} \left( \overline{D^{}_{\alpha\rm R}} \gamma^{}_\mu U^{}_{\beta\rm R} \right) \left( \ell^a_{\gamma\rm L} C {\rm i} D^\mu \ell^b_{\lambda\rm L} \right)$} & &  \\ $\Op^{(S)\alpha\beta\gamma\lambda}_{\overline{\ell}qddD} $ &  \makecell[c]{$ \displaystyle\frac{1}{2} \left( \Op^{\alpha\beta\gamma\lambda}_{\overline{\ell} qdd D} + \Op^{\alpha\beta\lambda\gamma}_{\overline{\ell} qdd D} \right)$ \\ with $\Op^{\alpha\beta\gamma\lambda}_{\overline{\ell} qdd D}  = \left( \overline{\ell^{}_{\alpha\rm L}} \gamma^{}_\mu Q^{}_{\beta\rm L} \right) \left( D^{}_{\gamma\rm R} C {\rm i} D^\mu D^{}_{\lambda\rm  R} \right)$} & &  \\
			\hline
			\hline
		\end{tabular}
	}
	\caption{A physical basis for dim-7 operators. Notations for operators and indices can be found in Ref.~\cite{Zhang:2023kvw}. }\label{tab:phyb}
\end{table}

The operator basis for dim-7 operators was first discussed in Ref.~\cite{Lehman:2014jma} and two redundant operators were removed from this basis later~\cite{Liao:2016hru}. However, there were still some redundancies in this basis due to nontrivial flavor relations among operators induced by equations of motion (EoMs)~\cite{Liao:2019tep}. Those redundancies were got rid of and a physical basis was put forward in Ref.~\cite{Liao:2019tep}. Here, we propose a new physical basis for dim-7 operators in Table~\ref{tab:phyb}~\cite{Zhang:2023kvw}. The difference between this basis and the one in Ref.~\cite{Liao:2019tep} is the way to determine the non-redundant degrees of freedom in operators $\Op^{}_{\overline{e}\ell\ell\ell H}$ and $\Op^{}_{\overline{\ell}dddH}$. We decompose those operators with flavor relations by means of SU(3) tensor decomposition since each fermion field can be regarded as the fundamental representation of a SU(3) flavor symmetry. For instance, $\Op^{}_{\overline{e}\ell\ell\ell H}$ can be decomposed into one totally symmetric, one totally anti-symmetric and two mixed-symmetric combinations, namely
\begin{eqnarray}
	\Op^{\alpha\beta\gamma\lambda}_{\overline{e}\ell\ell\ell H} &=& 	\Op^{(S)\alpha\beta\gamma\lambda}_{\overline{e}\ell\ell\ell H} + 	\Op^{(A)\alpha\beta\gamma\lambda}_{\overline{e}\ell\ell\ell H} + 	\Op^{(M)\alpha\beta\gamma\lambda}_{\overline{e}\ell\ell\ell H} + 	\Op^{(M^\prime)\alpha\beta\gamma\lambda}_{\overline{e}\ell\ell\ell H} \;.
\end{eqnarray}
The last combination $\Op^{(M^\prime)\alpha\beta\gamma\lambda}_{\overline{e}\ell\ell\ell H} = \left( \Op^{\alpha\beta\gamma\lambda}_{\overline{e}\ell\ell\ell H} + \Op^{\alpha\lambda\gamma\beta}_{\overline{e}\ell\ell\ell H} -  \Op^{\alpha\lambda\beta\gamma}_{\overline{e}\ell\ell\ell H} - \Op^{\alpha\gamma\beta\lambda}_{\overline{e}\ell\ell\ell H} \right)/3$ is automatically vanishing due to the flavor relation $\Op^{\alpha\beta\gamma\lambda}_{\overline{e}\ell\ell\ell H} + \Op^{\alpha\lambda\gamma\beta}_{\overline{e}\ell\ell\ell H} - \Op^{\alpha\lambda\beta\gamma}_{\overline{e}\ell\ell\ell H} - \Op^{\alpha\gamma\beta\lambda}_{\overline{e}\ell\ell\ell H} = 0$
whereas the other three explicitly given in Table~\ref{tab:phyb} are free from the flavor relation and unlike those in Ref.~\cite{Liao:2019tep}, their flavor indices get no constraints and run over all flavors. Therefore, the basis in Table~\ref{tab:phyb} is more suitable for calculations and to organize results in a compact form. Similarly, one can construct a Green's basis for dim-7 operators, where there are eight extra operators, i.e.,~\cite{Zhang:2023kvw}
\begin{eqnarray}
\Opr^{(S)\alpha\beta}_{\ell HD3} &=& \frac{1}{2} \left( \Opr^{\alpha\beta}_{\ell HD3} + \Opr^{\beta\alpha}_{\ell HD3} \right)  {\rm ~with~} \Opr^{\alpha\beta}_{\ell HD3} = \epsilon^{ad} \epsilon^{be} \left( \ell^a_{\alpha\rm L} C  \ell^b_{\beta\rm L} \right) D^\mu H^d D^{}_\mu H^e \;,
\nonumber
\\
\Opr^{(S)\alpha\beta}_{\ell HD4} &=& \frac{1}{2} \left( \Opr^{\alpha\beta}_{\ell HD4} + \Opr^{\beta\alpha}_{\ell HD4} \right)  {\rm ~with~} \Opr^{\alpha\beta}_{\ell HD4} = \epsilon^{ad} \epsilon^{be} \left( D^\mu \ell^a_{\alpha\rm L} C D^{}_\mu \ell^b_{\beta\rm L} \right) H^d H^e \;,
\nonumber
\\
\Opr^{\alpha\beta}_{\ell HD5} &=& \epsilon^{ab} \epsilon^{de} \left( \ell^a_{\alpha\rm L} C \sigma^{}_{\mu\nu} D^\mu \ell^b_{\beta\rm L} \right) H^d D^\nu H^e \;,
\nonumber
\\
\Opr^{(S)\alpha\beta}_{\ell HD6} &=& \frac{1}{2} \left( \Opr^{\alpha\beta}_{\ell HD6} + \Opr^{\beta\alpha}_{\ell HD6} \right) {\rm ~with~} \Opr^{\alpha\beta}_{\ell HD6} = \epsilon^{ad} \epsilon^{be} \left( D^\mu \ell^a_{\alpha\rm L} C \sigma^{}_{\mu\nu} D^\nu \ell^b_{\beta\rm L} \right) H^d H^e \;,
\nonumber
\\
\Opr^{\alpha\beta\gamma\lambda}_{\overline{d}\ell\ell D u} &=&\epsilon^{ab} \left( \overline{D^{}_{\alpha\rm R}} \ell^a_{\beta\rm L} \right) \left( \ell^b_{\gamma\rm L }  C \gamma^{}_\mu {\rm i}D^\mu U^{}_{\lambda\rm R } \right) \;,\quad \Opr^{\alpha\beta\gamma\lambda}_{\overline{d} D \ell\ell u} = \epsilon^{ab} \left( \overline{D^{}_{\alpha\rm R}} {\rm i}D^\mu \ell^a_{\beta\rm L} \right) \left( \ell^b_{\gamma\rm L} C \gamma^{}_\mu U^{}_{\lambda\rm R} \right) \;,
\nonumber
\\
\Opr^{\alpha\beta\gamma\lambda}_{\overline{\ell}dDqd} &=& \left( \overline{\ell^{}_{\alpha\rm L}} D^{}_{\beta\rm R} \right) \left( {\rm i}D^\mu Q^{}_{\gamma\rm L} C \gamma^{}_\mu D^{}_{\lambda\rm R} \right) \;,\quad  \Opr^{\alpha\beta\gamma\lambda}_{\overline{\ell} d q D d} = \left( \overline{\ell^{}_{\alpha\rm L}} D^{}_{\beta\rm R} \right) \left( Q^{}_{\gamma\rm L} C \gamma^{}_\mu {\rm i}D^\mu D^{}_{\lambda\rm R} \right) \;,
\end{eqnarray}
apart from the operators in Table~\ref{tab:phyb} with $\Op^{(S)}_{\ell HD1}$, $\Op^{(S)}_{\ell HD2}$, $\Op^{(S)}_{\overline{e}dddD}$, $\Op^{(S)}_{\overline{d}u\ell\ell D}$ and $\Op^{(S)}_{\overline{\ell}qddD}$ replaced by $\Op^{}_{\ell HD1}$, $\Op^{}_{\ell HD2}$, $\Op^{}_{\overline{e}dddD}$, $\Op^{}_{\overline{d}u\ell\ell D}$ and $\Op^{}_{\overline{\ell}qddD}$. By mean of the SM fields' EoMs, one can obtain the reduction relations to convert operators in the Green's basis to those in the physical basis, e.g.,
\begin{eqnarray}
	C^{\alpha\beta}_{\ell eHD} &=& G^{\alpha\beta}_{\ell eHD} +  \frac{1}{2} \left( G^\dagger_5 \right)^{\alpha\gamma} \left[ G^{\gamma\beta}_{eHD2} - G^{\gamma\beta}_{eHD4} - 2 G^{\gamma\lambda}_{\ell D} \left( Y^{}_l \right)^{}_{\lambda\beta} \right]   - \frac{1}{4}  \left( G^{\alpha\gamma}_{\ell HD2} - G^{\gamma\alpha}_{\ell HD2} \right)\left( Y^{}_l \right)^{}_{\gamma \beta} 
	\nonumber
	\\
	&& + \left( \rmi G^{\alpha\gamma}_{\ell HD5} + \rmi G^{(S)\alpha\gamma}_{\ell HD6} - G^{(S)\alpha\gamma}_{\ell H D 4} \right) \left( Y^{}_l \right)^{}_{\gamma\beta} \;,
\end{eqnarray}
where some redundant dim-6 operators are involved due to  the existence of $\Op^{(5)}$. More details for basis constructions for dim-7 operators and the full reduction relations can be found in Ref.~\cite{Zhang:2023kvw}.

\section{RGEs of Dim-5 and Dim-7 Operators in the SMEFT}

Starting with the physical basis in Table~\ref{tab:phyb}, we calculate a set of 1PI diagrams to extract counterterms in the Green's basis with the modified minimal subtraction scheme, and then with the help of reduction relations, we obtain all counterterms in the physical basis, from which one can derive RGEs for Wilson coefficients of all operators. Since the calculations and results are pretty lengthy (see Refs.~\cite{Zhang:2023kvw,Zhang:2023ndw}), we only show two examples for the coefficients of $\Op^{(5)}$ and $\Op^{(S)}_{\ell H}$ here:
\begin{eqnarray} \label{eq:rge-wein5}
	\hspace{-3cm}\befun^{\alpha\beta}_5 &=& \frac{1}{2} \left( -3g^2_2 + 4\lambda +2T \right) C^{\alpha\beta}_5 - \frac{3}{2}  \left( Y^{}_l Y^\dagger_l C^{}_5 \right)^{\alpha\beta} + m^2 \left( 8C^{}_{H\square} - C^{}_{HD} \right)  C^{\alpha\beta}_5 + m^2 \left\{ 8 C^{(S)\ast \alpha\beta}_{\ell H}  \right.
	\nonumber
	\\
	&& + \frac{3}{2} g^2_2 \left( 2 C^{(S)\ast \alpha\beta}_{\ell HD1} + C^{(S)\ast \alpha\beta}_{\ell HD2}  \right) +  \left( Y^{}_l Y^\dagger_l  C^{(S)\dagger}_{\ell HD1} \right)^{\alpha\beta} - \frac{1}{2} \left( Y^{}_l Y^\dagger_l  C^{(S)\dagger}_{\ell HD2} \right)^{\alpha\beta}   + 2\left( Y^{}_l C^\dagger_{\ell e HD} \right)^{\alpha\beta}   
	\nonumber
	\\
	&&  - \left. \left( Y^\dagger_l \right)^{}_{\gamma\lambda} \left( 3 C^{(S)\ast\gamma\lambda\alpha\beta}_{\overline{e}\ell\ell\ell H}  + 2 C^{(M)\ast\gamma\lambda\alpha\beta}_{\overline{e}\ell\ell\ell H}  \right) - 3 \left( Y^{\dagger}_{\rm d} \right)^{}_{\gamma\lambda} C^{\ast\gamma\alpha\lambda\beta}_{\overline{d}\ell q\ell H1}  + 6 \left( Y^{}_{\rm u} \right)^{}_{\lambda\gamma}  C^{\ast\lambda\gamma\alpha\beta}_{\overline{q}u\ell\ell H}   \right\}  + \alpha \leftrightarrow \beta \;,
	\nonumber
	\\
	\hspace{-4cm}\befun^{(S)\alpha\beta}_{\ell H} &=&  \frac{1}{2} \tr{C^{}_5 C^\dagger_5} C^{\ast \alpha\beta}_5 +  \frac{5}{4} \left( C^\dagger_5 C^{}_5 C^\dagger_5 \right)^{\alpha\beta}   +  C^{\ast \alpha\beta}_5 \left\{  - \frac{3}{4} \left( g^2_1 - g^2_2 + 4\lambda \right) C^{}_{HD} + \left( 16 \lambda - \frac{5}{3} g^2_2 \right) C^{}_{H \square}  \right. 
	\nonumber
	\\
	&& -3 C^{}_H - 3g^2_2 C^{}_{HW} + \frac{3}{2} \rmi \left( g^2_1 C^{}_{H\widetilde{B}} + 3g^2_2 C^{}_{H\widetilde{W}} + g^{}_1g^{}_2 C^{}_{H\widetilde{W}B} \right) - {\rm Tr} \left[ 2g^2_2 \left(  C^{(3)}_{Hq} + \frac{1}{3} C^{(3)}_{H\ell} \right)+ C^{}_{eH} Y^\dagger_l  \right.
	\nonumber
	\\
	&& + \left.\left.\hspace*{-0.15cm} 3C^{}_{dH} Y^\dagger_{\rm d} + 3Y^{}_{\rm u} C^\dagger_{uH} - 2 \left( Y^\dagger_l C^{(3)}_{H\ell} Y^{}_l + 3 Y^\dagger_{\rm d} C^{(3)}_{Hq} Y^{}_{\rm d} + 3 Y^\dagger_{\rm u} C^{(3)}_{Hq} Y^{}_{\rm u} \right)  + 3  \left( Y^{}_{\rm u} C^{}_{Hud} Y^\dagger_{\rm d} + Y^{}_{\rm d} C^\dagger_{Hud} Y^\dagger_{\rm u} \right) \vphantom{\frac{1}{3}}\right] \right\} 
	\nonumber
	\\
	&& - 3 g^{}_2 \left( C^\dagger_5 Y^{}_l C^\dagger_{eW} \right)^{\alpha\beta} + \frac{3}{2} \left( g^2_1 + g^2_2 \right) \left[ \left( C^\dagger_5 C^{(3)}_{H\ell} \right)^{\alpha\beta} - \left( C^\dagger_5 C^{(1)}_{H\ell} \right)^{\alpha\beta}  \right]  + \frac{1}{2}  \left( C^\dagger_5 Y^{}_l C^\dagger_{eH} \right)^{\alpha\beta} 
	\nonumber
	\\
	&&  +  \left( C^\dagger_5 C^{}_{eH} Y^\dagger_l  \right)^{\alpha\beta} - 3 \left( C^\dagger_5 Y^{}_l Y^\dagger_l C^{(3)}_{H\ell} \right)^{\alpha\beta}  - \frac{1}{4} \left( 3g^2_1 + 15g^2_2 - 80 \lambda - 8 T \right) C^{(S)\alpha\beta}_{\ell H} - \frac{3}{2} \left( C^{(S)}_{\ell H} Y^{}_l Y^\dagger_l  \right)^{\alpha\beta} 
	\nonumber
	\\
	&& + \left( 2\lambda - \frac{3}{2} g^2_2 \right) \left( C^{}_{\ell e HD} Y^\dagger_l \right)^{\alpha\beta} + \left( C^{}_{\ell e HD} Y^\dagger_l Y^{}_l Y^\dagger_l \right)^{\alpha\beta}  - \frac{3}{4} g^2_2  \left(  g^2_2 - 4 \lambda \right) C^{(S)\alpha\beta}_{\ell HD1} + \lambda \left( C^{(S)}_{\ell HD1} Y^{}_l  Y^\dagger_l \right)^{\alpha\beta} 
	\nonumber
	\\
	&& - \left( C^{(S)}_{\ell HD1} Y^{}_l  Y^\dagger_l Y^{}_l Y^\dagger_l  \right)^{\alpha\beta} - \frac{3}{8} \left( g^4_1 + 2 g^2_1 g^2_2 + 3g^4_2 - 4g^2_2 \lambda \right) C^{(S)\alpha\beta}_{\ell HD2}  - \frac{1}{2} \lambda \left( C^{(S)}_{\ell HD2} Y^{}_l  Y^\dagger_l \right)^{\alpha\beta} 
	\nonumber
	\\
	&& - \left( C^{(S)}_{\ell HD2} Y^{}_l  Y^\dagger_l Y^{}_l Y^\dagger_l  \right)^{\alpha\beta} - 3 g^3_2 C^{\alpha\beta}_{\ell HW} - 6 g^{}_2 \left( C^{}_{\ell HW} Y^{}_l Y^\dagger_l \right)^{\alpha\beta} - 3 C^{(S)\gamma\lambda\alpha\beta}_{\overline{e}\ell\ell\ell H} \left[ \lambda \left( Y^{}_l \right)^{}_{\lambda\gamma} - \left( Y^{}_l Y^\dagger_l Y^{}_l \right)^{}_{\lambda\gamma}  \right]  
	\nonumber
	\\
	&&   - 2C^{(M)\gamma\lambda\alpha\beta}_{\overline{e}\ell\ell\ell H}   \left[  \lambda  \left( Y^{}_l \right)^{}_{\lambda\gamma} - \left( Y^{}_l Y^\dagger_l Y^{}_l \right)^{}_{\lambda\gamma} \right] - 3 C^{\gamma\alpha\lambda\beta}_{\overline{d}\ell q\ell H1} \left[ \lambda  \left( Y^{}_{\rm d} \right)^{}_{\lambda\gamma} -  \left( Y^{}_{\rm d} Y^\dagger_{\rm d} Y^{}_{\rm d} \right)^{}_{\lambda\gamma}  \right]
	\nonumber
	\\
	&&  + 6 C^{\gamma\lambda\alpha\beta}_{\overline{q}u\ell\ell H} \left[ \lambda \left( Y^\dagger_{\rm u} \right)^{}_{\lambda\gamma} - \left( Y^\dagger_{\rm u} Y^{}_{\rm u} Y^\dagger_{\rm u} \right)^{}_{\lambda\gamma} \right] + \alpha \leftrightarrow \beta \;,
\end{eqnarray}
which satisfy the general form in Eq.~\eqref{eq:rge-form}. The nonrenormalization theorem~\cite{Cheung:2015aba} can predict the zero entries in the anomalous dimension matrix for mixing among the same dimensional operators. Based on our results, we show the anomalous dimension matrix $\gamma^{(7,7)}_{ij}$ for baryon-number-conserving operators in Table~\ref{tab:adm}, where the zero entries in light grey cells and the non-vanishing entries in dark grey cells having Yukawa couplings of nonholomorphic forms are fully consistent with the nonrenormalization theorem. One may also check $\gamma^{(7,7)}_{ij}$ for baryon-number-violating operators and find they are coincident with the nonrenormalization theorem as well~\cite{Zhang:2023ndw}. Moreover, our results for mixing among different dimensional operators, e.g., $\gamma^{(7,6)}_{ij}$, may give an insight into a non-linear version of the theorem. One may refer to Refs.~\cite{Zhang:2023kvw,Zhang:2023ndw} for more discussions about the results.

\begin{table}
	\centering
	\renewcommand\arraystretch{1.3}
	\resizebox{0.9\textwidth}{!}{
		\begin{tabular}{c|cc|c|cc|ccc|c|cc|c}
			\hline\hline
			\diagbox{ $C^{}_i$ $(w_i,\overline{w}_i)$}{$\gamma^{}_{ij}$}{\makecell[c]{ $C^{}_j$\\ $(w_j,\overline{w}_j)$} } & \makecell[c]{ $C^{(S)}_{\ell HD1}$ \\ $(3,5)$} & \makecell[c]{ $C^{(S)}_{\ell HD2}$ \\ $(3,5)$}  & \makecell[c]{ $C^{(S)}_{\overline{d}u\ell\ell D}$ \\ $(3,5)$} & \makecell[c]{ $C^{(A)}_{\ell HB}$  \\ $(3,7)$} & \makecell[c]{$C^{}_{\ell HW}$ \\ $(3,7)$} & \makecell[c]{ $C^{(S,A,M)}_{\overline{e}\ell\ell\ell H}$  \\ $(3,7)$}  & \makecell[c]{ $C^{}_{\overline{d}\ell q\ell H1}$ \\ $(3,7)$} & \makecell[c]{ $C^{}_{\overline{d}\ell q\ell H2}$ \\ $(3,7)$} & \makecell[c]{ $C^{}_{\ell eHD}$ \\ $(5,5)$}  & \makecell[c]{ $C^{}_{\overline{d}\ell ueH}$ \\ $(5,5)$} & \makecell[c]{ $C^{}_{\overline{q}u\ell\ell H}$ \\ $(5,5)$} & \makecell[c]{ $C^{(S)}_{\ell H}$\\ $(5,7)$} \\
			\hline
			$C^{(S)}_{\ell HD1}$ $(3,5)$  & $g^2,y^2$ & $g^2,\lambda$ & $y^2$ & \cellcolor{gray!30}{0}  & \cellcolor{gray!30}{0}  & \cellcolor{gray!30}{0}  & \cellcolor{gray!30}{0}  & \cellcolor{gray!30}{0} & \cellcolor{gray!30}{0}  & \cellcolor{gray!30}{0}  & \cellcolor{gray!30}{0} & \cellcolor{gray!30}{0}  \\
			$C^{(S)}_{\ell HD2}$ $(3,5)$  & $g^2,y^2$ & $g^2,y^2, \lambda$ & 0  & \cellcolor{gray!30}{0}  & \cellcolor{gray!30}{0}  & \cellcolor{gray!30}{0}  & \cellcolor{gray!30}{0}  & \cellcolor{gray!30}{0} & \cellcolor{gray!30}{0}  & \cellcolor{gray!30}{0} & \cellcolor{gray!30}{0}  & \cellcolor{gray!30}{0} \\
			\hline
			$C^{(S)}_{\overline{d}u\ell\ell D}$ $(3,5)$ &  $y^2$ & $y^2$ &  $g^2, y^2$ &  \cellcolor{gray!30}{0} &  \cellcolor{gray!30}{0} & \cellcolor{gray!30}{0} &  \cellcolor{gray!30}{0} &  \cellcolor{gray!30}{0} &  \cellcolor{gray!30}{0}  &  \cellcolor{gray!30}{0} &  \cellcolor{gray!30}{0} &  \cellcolor{gray!30}{0} \\
			\hline
			$C^{(A)}_{\ell HB}$  $(3,7)$ & $g y^2$ & $gy^2$  & 0 & $g^2, y^2, \lambda$ & $g^2$  & $gy$ & $gy$ & 0 & \cellcolor{gray!30}{0} & \cellcolor{gray!30}{0}  & \cellcolor{gray!30}{0} & \cellcolor{gray!30}{0} \\
			$C^{}_{\ell HW}$ $(3,7)$ & $g^3, g y^2$ & $g^3, gy^2$ & 0 & $g^2$ & $g^2, y^2, \lambda$  & $gy$ & $gy$ & $gy$  & \cellcolor{gray!30}{0} & \cellcolor{gray!30}{0}  & \cellcolor{gray!30}{0}  & \cellcolor{gray!30}{0} \\
			\hline
			$C^{(S,A,M)}_{\overline{e}\ell\ell\ell H}$  $(3,7)$ & $g^2y, y^3$ & $g^2y, y^3$ & 0 & $gy$ & $gy$ &  $g^2, y^2$ & $y^2$ & $y^2$ & \cellcolor{gray!30}{0} & \cellcolor{gray!30}{0} & \cellcolor{gray!70}{$\overline{y}^2$} & \cellcolor{gray!30}{0} \\
			$C^{}_{\overline{d}\ell q\ell H1}$  $(3,7)$  & $g^2y, y^3$ & $g^2y, y^3$ & $g^2y,y^3$ & $gy$ & $gy$ & $ y^2$ & $g^2, y^2$ & $g^2, y^2$ & \cellcolor{gray!30}{0} & \cellcolor{gray!70}{$\overline{y}^2$} & \cellcolor{gray!70}{$\overline{y}^2$} & \cellcolor{gray!30}{0} \\
			$C^{}_{\overline{d}\ell q\ell H2}$  $(3,7)$ & $g^2y, y^3$ & $g^2y, y^3$ & $g^2y,y^3$ & $gy$ & $gy$ & $ y^2$ & $g^2, y^2$ & $g^2, y^2$ & \cellcolor{gray!30}{0} & \cellcolor{gray!70}{$\overline{y}^2$} & \cellcolor{gray!70}{$\overline{y}^2$} & \cellcolor{gray!30}{0} \\
			\hline
			$C^{}_{\ell eHD}$ $(5,5)$  &$g^2y, y^3$ & $g^2y, \lambda y, y^3$ & 0 &  \cellcolor{gray!30}{0}  &  \cellcolor{gray!30}{0} & \cellcolor{gray!30}{0}  &  \cellcolor{gray!30}{0} &  \cellcolor{gray!30}{0}  & $g^2,y^2,\lambda$  & $y^2$ & 0 &    \cellcolor{gray!30}{0} \\
			\hline
			$C^{}_{\overline{d}\ell ueH}$ $(5,5)$  & $y^3$ & $y^3$ & $g^2y, y^3$ & \cellcolor{gray!30}{0}  & \cellcolor{gray!30}{0} & \cellcolor{gray!30}{0}  & \cellcolor{gray!70}{$y^2$} & \cellcolor{gray!70}{$y^2$}  & $y^2$ & $g^2,y^2$ & $y^2$  & \cellcolor{gray!30}{0}  \\
			$C^{}_{\overline{q}u\ell\ell H}$ $(5,5)$ & $g^2y, y^3$ & $g^2y, y^3$ & $g^2y, y^3$ & \cellcolor{gray!30}{0}  & \cellcolor{gray!30}{0} & \cellcolor{gray!70}{$y^2$}  & \cellcolor{gray!70}{$y^2$} & \cellcolor{gray!70}{$y^2$}  & 0 & $y^2$ & $g^2, y^2$ & \cellcolor{gray!30}{0}  \\
			\hline
			$C^{(S)}_{\ell H}$ $(5,7)$ & $g^4,\lambda g^2, \lambda y^2, y^4$ & $g^4,\lambda g^2, \lambda y^2, y^4$ & 0 & 0 & $g^3,gy^2$ & $\lambda y, y^3$ & $\lambda y, y^3$ & 0 & $\lambda y, g^2 y , y^3$ & 0 & $\lambda y, y^3$ & $g^2,y^2,\lambda$ \\
			\hline\hline
		\end{tabular}
	}
	\caption{The structure of the one-loop anomalous dimension matrix $\gamma^{(7,7)}_{ij}$ for Wilson coefficients of dim-7 baryon-number-conserving operators. $w$ and $\overline{w}$ are holomorphic and antiholomorphic weights of operators.}
	\label{tab:adm}
\end{table}

\section{Summary}

We have proposed a new physical basis and a Green's basis for dim-7 operators in the SMEFT, where there are no constraints on operators' flavor indices and they can run over all flavors. Therefore, those bases are suitable for matching and derivation of RGEs, and can also keep results in a compact form. Based on those two bases and the reduction relations among them, we have derived the complete one-loop RGEs of dim-5 and dim-7 operators up to $\mathcal{O} \left( \Lambda^{-3} \right)$ in the SMEFT. With those obtained results, one can discuss full RG running effects on some appealing lepton- or baryon-number-violating observables or processes up to $\mathcal{O}\left( \Lambda^{-3} \right) $ in the SMEFT, such as neutrino masses, neutrinoless double beta decay, meson and nucleon decays (see, e.g., Refs.~\cite{Isidori:2023pyp,Chala:2021juk,Liao:2016hru,Liao:2019tep,Zhang:2023kvw,Zhang:2023ndw} and references therein).

{\it This work is supported by the Alexander von Humboldt Foundation.}

\end{document}